\documentclass[11pt]{article}
\usepackage{amssymb,euscript,latexsym,amsmath}
\usepackage{graphicx}

\makeatletter
\@addtoreset{figure}{section}
\def\thefigure{\thesection.\@arabic\c@figure}
\def\fps@figure{h, t}
\@addtoreset{table}{section}
\def\thetable{\thesection.\@arabic\c@table}
\def\fps@table{h, t}
\@addtoreset{equation}{section}

\makeatother

\begin{document}
\pagestyle{empty}

\title{Control of Squeezed States}

\author{Anthony M. Bloch\thanks{Research partially supported by the
National Science Foundation and the Air Force Office of Scientific
Research}
\\Department of Mathematics
\\ University of Michigan
\\ Ann Arbor, MI 48109
\\ {\small abloch@math.lsa.umich.edu}
\and
Alberto G. Rojo\thanks{Research partially supported by the
National Science Foundation.}
\\ Dept. of Physics
\\ University of Michigan
\\ Ann Arbor, MI 48109
\\{\small rojoa@umich.edu}
}
\maketitle

\begin{abstract}
In this paper we consider the classical and quantum control of squeezed states
of harmonic oscillators. This provides a method for reducing
noise below the quantum limit and provides an example of the control
of under-actuated systems in the stochastic and quantum context. 
We consider also the interaction of a squeezed quantum oscillator with
an external heat bath.

\end{abstract}


\section{Introduction}
In this paper we consider the problem of squeezing of harmonic oscillators
from the point of view of control theory. Squeezing has been suggested
as a method for reducing noise in quantum systems below the standard 
quantum limit. This can be achieved by using laser pulses and in that
sense may be viewed as a quantum control problem, although the
classical squeezing problem is also of interest. In the latter case one
is interested in reducing noise induced by random perturbations.

The quantum control problem has been of great interest recently,
see for example Brockett and Khaneja [1999],
Lloyd [1996] and Warren et. al. [1993] and references therein.

Here we consider squeezing as a control problem in both the classical
and quantum setting. In the classical case we consider a system subject
to thermal noise while in the quantum case we consider a system
at zero temperature and in the 
presence of noise. In both cases the control is given by an external
electromagnetic field and enters the control equations multiplicatively. 
In this sense the setting is similar to the NMR control problems
analyzed by Brockett and Khaneja.

A key feature of squeezing is that it results in a redistribution
of uncertainty between observables.

In this paper we consider a model for phonon squeezing in solids following
the work of Garret at. al. [1997], but one can equally well consider
the case of photons in quantum optics.
The control is via a single pulse on a large 
ensemble of oscillators and this sense we are considering under-actuated
control systems in both the classical and quantum case.

We also model the effect of dissipation on the classical system and the 
effect of coupling to a heat bath in the quantum setting. This causes
the squeezing effect to gradually moderate.

\section {The Control Setting}
In  the classical setting we consider bilinear control systems of the form
\begin{equation}
\dot{x^i}=\sum_{j=1}^na_{ij}x^j+\sum_{j=1}^mb_{ij}(x)u^j+
\sum_{j=1}^rg_{ij}\dot{w^k}
\label{linmultform}
\end{equation}
where $a_{ij}$ is a constant matrix (i.e. the 
free dynamics is linear), $b_{ij}$ is linear in $x$, i.e. the control $u^j$
enters bilinearly, $\dot{w^k}$ is white noise and the state
space is $\mathbb{R}^n$. 

In the quantum context we want to consider a similar equation
but defined on an appropriate Hilbert space:

\begin{equation}
ih\frac{\partial\psi}{\partial t}
=H_s\psi+\sum_ju_jH^j\psi+\sum_j\dot{w^j}H_j\psi
\end{equation}
where $H_s$ is the Schroedinger operator, the $H_j$ are (linear) 
input operators, $u^j$ are functions of time, and the $\psi$ is a
vector in the Hilbert space.

\section{Classical Squeezing of the Harmonic Oscillator}

In this section we consider classical squeezing of a set of identical
coupled harmonic oscillators. Denote the position of each oscillator
by $u^i$.

The Hamiltonian for the system
is of the form:

\begin{equation}
H=\sum_i\frac{p_i^2}{2}+\sum_i\frac{\omega_i^2}{2}(q^i-q^{i+1})^2
\end{equation}
where the oscillators are assumed to have unit mass and $p_i=\dot{u^i}$.

In order to analyze the system we decompose it into its normal modes.
Denoting the normal mode coordinates by $Q^i$ we thus obtain a system of
uncoupled harmonic oscillator equations of the form
$\ddot{Q^i}+\Omega_i^2Q^i=0$.

The main control mechanism we consider here is squeezing by pulses.
In this case each oscillator is forced by a pulse at time $t=0$ which
is proportional to its displacement, i.e. we have equations of the form:

\begin{equation}
\ddot{Q^i}+\Omega_i^2Q^i=2\lambda Q^i\delta(t)
\end{equation}
where $\delta(t)$ is the Dirac delta function and $\lambda$ is a constant
which is proportional to the frequency $\Omega$. 

Thus we obtain
\begin{equation}
\dot{Q}^i(0^+)=\dot{Q}^i(0^-)+2\lambda Q^i(0)\,.
\end{equation}

Thus, if one considers the system subject to white noise,
\begin{equation}
\ddot{Q^i}+\Omega_i^2Q^i=2\lambda Q^i\delta(t)+\alpha\dot{w}^i\,,
\end{equation}

one sees that while one starts with a spherical equilibrium distribution
which is invariant in time, after the pulse one has an elliptical
distribution which rotates in time at twice the harmonic frequency
(by the $\mathbb{Z}_2$ symmetry of the ellipse). (A precise
analysis is  given below in the course of our treatment of 
the quantum mechanical case.)
Noise reduction is then
achieved by viewing the system ``stroboscopically'' when the noise is low.

Actually the above is an idealization: in actuality the oscillator
should be viewed as in equilibrium with a heat bath which dissipates
energy. In the classical setting one can model this by simple
linear dissipation (in the quantum setting one has to introduce
a heat bath -- see below).

Thus we have a system of the form
\begin{equation}
\ddot{Q^i}+\Omega_i^2Q^i=-\eta_i\dot{Q}^i+U_i(t)+\alpha\dot{w}^i
\end{equation}
where $\eta_i$ is a dissipation constant and $U_i(t)$ is the 
control which we can choose to be a single pulse or a sequence of pulses.
Depending on the dissipation strength an initial squeezing effect will
decay away and we need a continual sequences of pulse to keep the 
system in a squeezed state.

It is worthwhile remarking on the how the control enters in
our setting: the control is a single pulse applied overall (and in this
sense the system is under-actuated) while
 the effect on each (normal mode) of oscillation is to apply a pulse
 proportional to displacement (minus the mean displacement which is of course
 zero for each oscillator). This is effected by the type of interaction
 of the oscillators with the field that the pulse induces. We note also
that in the full nonlinear setting the mean displacement may not be zero
and must be taken into account.
               
We shall return to the classical squeezing of oscillator by pulses, and
in particular a computation of mean square displacement,
after a discussion of the quantum case below.

We note also that parametric resonance control can achieve
similar squeezing effects in the classical case. 
In this case we consider oscillator motion in the presence of 
a modulating drive:
\begin{equation}
\ddot{Q}-\omega^2(1+\epsilon \cos2\omega t)Q=0
\end{equation}
where $\epsilon$ parameterizes the strength of the drive. We omit
details of this approach here.

\section{Squeezing of the Quantum Harmonic Oscillator}
We now turn to 
the quantum setting.

Consider the following  Hamiltonian
\begin{equation}
H= {P^2\over 2m} + {m\omega^2\over 2} Q^2 + \lambda \delta(t) Q^2,
\end{equation}
which reflects an impulsive change in the spring constant and where
$\omega=\sqrt{K/m}$, $K$ being the original spring constant.

The variables $P$ and $Q$, which are operators in the quantum case,
 obey canonical commutation rules $[P,Q]=i\hbar$. 
We can rewrite the above Hamiltonian in terms of creation operators $a$ and $a^{\dagger}$
defined
through
\begin{equation}
Q=\sqrt{\hbar \over 2 m \omega}(a + a^{\dagger}),
\,\,P=i\sqrt{\hbar  m \omega\over 2}( a^{\dagger}- a),
\end{equation}
with $[a,a^{\dagger}]=1$.
Written in terms of the new variables, the Hamiltonian is
\begin{equation}
H=\hbar \omega (a^{\dagger}a +1/2) + \lambda \delta(t) (a+a^{\dagger})^2.
\end{equation}
The ground state of the system, for $t\neq 0$, $|0\rangle$, corresponds to
 the vacuum of $a$, 
($a|0\rangle =0$), and the excited states are of the form $(a^{\dagger})^2|0\rangle$.

We now want to study the behavior of the system at $t>0$, given that the system is in
its ground state at $t<0$. The wave function at $t=0^+$ is of the form
$
|\psi(t=0^+)\rangle = {\rm exp}(-i\lambda Q^2)|0\rangle,
$
and for longer  times the system evolves with the ``unperturbed" Hamiltonian:
$
|\psi(t>0)\rangle = {\rm exp}(-iH_0 t)e^{-i\lambda Q^2}|0\rangle.
$
Our first quantity of interest is $\langle\psi(t)| Q^2|\psi(t)\rangle\equiv 
\langle Q^2(t)\rangle$. Let us 
compute it using the general method of coherent states. We find

\begin{equation}
\langle Q^2(t)\rangle=\langle 0|e^{i\lambda Q^2}(ae^{-i\omega t}
+ a^{\dagger}e^{-i\omega t})^2e^{-i\lambda Q^2}|0\rangle,
\label{x2t}
\end{equation}
where we have used the fact that $e^{iH_0 t}ae^{-iH_0 t}=ae^{-i\omega t}$, which 
states that $a^{\dagger}$ and $a$ respectively destroy and create
 eigenstates of
$H_0$,  and where $Q$ is defined in units of 
$\sqrt{\hbar/(2 m \omega)}$.

Now we introduce a basis of coherent states $|z\rangle$, which satisfy
$
a|z\rangle = z |z\rangle, \;\;\;\;\; \langle z| a^{\dagger}=\langle z| z^*, 
$
and form an overcomplete set of states:
\begin{equation}
1={1\over 2 \pi i} \int dz \; dz^* e^{-zz^*} |z\rangle\langle z|.
\label{complete}
\end{equation}

Inserting (\ref{complete}) in (\ref{x2t}) we find 
\begin{eqnarray*}
&&\langle Q^2(t)\rangle={1\over 2 \pi i}
 \int \int dz \; dz^* e^{-zz^*}\nonumber\\
&&(z ^2 e^{-2i\omega t}
 + z^{*2} e^{2i\omega t} +2zz^* -1) |\langle 0|e^{i\lambda x^2}|z\rangle |^2.
 \end{eqnarray*}
 In order to evaluate the last term we need the position representation of
 the ground state (note that at this point $Q$ is a real number) 
 \begin{equation}
 \langle 0|Q\rangle ={1\over \pi^{1\over 4}} e^{-Q^2/2}
 \end{equation}
 and that of the coherent state
 \begin{equation}
 \langle Q|z\rangle ={1\over \pi^{1\over 4}} e^{-Q^2/2 + \sqrt{2} zQ - z^2/2}.
 \end{equation}
 
 A simple integration gives
 \begin{eqnarray}
 \langle 0|e^{i\lambda Q^2}|z\rangle &=& \int dx \langle 0|Q\rangle \langle Q|z
 \rangle e^{i\lambda Q^2}\\
 &=& {1\over \sqrt{1-i\lambda} } e^{i\lambda z^2/2(1-i\lambda)}.
 \end{eqnarray}
 
 Changing to the variables $z=u+iv$ we have
 \begin{equation} 
 e^{-zz^*}|\langle 0|e^{i\lambda Q^2}|z\rangle |^2
 ={1\over \sqrt {1+\lambda ^2}} e^{-[v^2+(2\lambda^2+1)u^2 +2\lambda uv]/(1+\lambda^2)},
 \end{equation}
 and 
 \begin{eqnarray}
 &&\langle Q^2(t)\rangle={4\over\pi\sqrt{ 1+\lambda^2}}
 \int_{-\infty}^{\infty} du \int_{-\infty}^{\infty} dv\nonumber\\
 &&\left(u^2\cos^2\omega t +v^2 \sin ^2 \omega t
 +uv\sin 2\omega t - {1\over 4}
 \right)
 \nonumber
 \\ & & \times e^{-[v^2+(2\lambda^2+1)u^2 +2\lambda uv]/(1+\lambda^2)}\\
 &=&1+4\lambda ^2 \sin ^2 \omega t +2\lambda \sin 2\omega t
 \end{eqnarray}
 
 It is interesting to compare this with an ensemble of classical oscillators
 with initial conditions taken from a heat bath. For simplicity let us take
 $\omega = m=k_B=T=1$ ($k_B$ is Boltzman's constant).
 An arbitrary oscillator will evolve as 
 $
 Q(t) =u \cos t + v \sin t, 
 $
 with $u$ and $v$ its initial position and velocity. If a pulse is applied
 at $t=0$ of the form treated above:
 $Q(t) =u \cos t + (v+2\lambda u) \sin t. 
 $
 Now let us average over initial conditions taken from a measure given
 by (a thermal bath)
 \begin{eqnarray}
 \langle Q^2(t)\rangle&\sim&\int du \;
 dv [u \cos t + (v+2\lambda u) \sin t]^2 e^{-(u^2 +v^2)}
\nonumber \\
 &=&1+4\lambda^2 \sin ^2 t + 2\lambda \sin 2t.
 \label{nodamp}
 \end{eqnarray}
 
 It is interesting to note that the two expressions for, respectively,
 the   quantum oscillator at zero temperature and the classical oscillator
 at finite temperature, are exactly the same.
 The general time dependence of the variance for a squeezed harmonic oscillator 
with frequency $\omega$ can thus be written in
 the following form
\begin{equation}
 \langle [Q(t)]^2\rangle = {\epsilon _0\over K} \left[1+ 
 \left( {2\lambda  \over \omega } \right) \sin 2 \omega t
 + \left( { 2\lambda \over \omega } \right)^2 \sin ^2 \omega t \right]
\end{equation}
with $\epsilon _0=\hbar \omega/2$ for the quantum case and $\epsilon _0=k_BT$ for the 
classical oscillator at a temperature $T$.

The method of coherent states presented above has the advantage of being suitable
for calculating other quantities. For example, if the oscillators are atoms within
a solid, the scattering amplitude for an X-ray is  decreased by a factor (called
the Debye-Waller factor -- see Ziman [1972])
$\sim \langle \exp {ik Q(t)} \rangle$, with $k$ the wave-vector of the 
X-ray.  We now  ask ourselves what is the time evolution of the Debye-Waller
factor for a squeezed phonon. This means that we need to compute the following
expression
\begin{eqnarray}
&&I(\lambda,t)= 
\langle 0|e^{i\lambda Q^2}e^{(ae^{-i\omega t}
+ a^{\dagger}e^{-i\omega t})}e^{-i\lambda Q^2}|0\rangle
\nonumber\\
&&= {1\over \sqrt{e}}{1\over \sqrt{1+\lambda ^2}}{1\over \pi}
\int du \; dv\;\nonumber\\
&& e^{2u\cos \omega t +2 v\sin \omega t - 
{-[v^2+(2\lambda^2+1)u^2 +2\lambda uv]\over (1+\lambda^2)}}\nonumber\\
&=&e^{1+4\lambda ^2 \sin ^2 \omega t +2\lambda \sin 2\omega t}.
\label{DWt}
\end{eqnarray}

For the Debye-Waller factor, we obtain the following time dependence
\begin{equation}
 \langle e^{ik Q(t)} \rangle = e^{-k^2\langle { Q^2(t)} \rangle}
 \end{equation}

Measurement of the Debye-Waller factor may provide a practical method
of detecting the squeezing phenomenon experimentally.

\section{Squeezing and dissipation}
In this section we consider the squeezing of a quantum oscillator
coupled to a an infinite number of oscillators representing a ``heat" bath.
We show that this causes a decay in the squeezing
oscillation for small time 
and true damping in the limit of a continuum of oscillators. This 
damping effect of the heat bath is similar to that analyzed classically
in Lamb [1900], Komech [1995], Sofer and Weinstein [1999]
and Hagerty, Bloch and Weinstein [1999]. 
We stress that we are considering a zero temperature case, and 
the damping effects appear due to a) the coupling of a single variable
with a continuum of variables and b) an ``asymmetry" in the initial conditions.
The applied
pulse on the oscillator generates outgoing waves on the continuum system which
in turn gives rise to a positive damping (for a detailed discussion of 
negative versus positive damping see Keller and Bonilla [1986]).

We start with a general formulation, and at the end of this section discuss 
a specific continuum example.

The Hamiltonian of the system consists of three parts: $H_0$ describing
the original oscillator:
   \begin{equation}
   H_0= {p_0^2\over 2m} + {m\omega_0^2\over 2} q_0^2\, ,
   \end{equation}
   the Hamiltonian $H_e$ of the environment:
    \begin{equation}
   H_e= \sum _\alpha\left[ {p_\alpha^2\over 2m} + {m\omega_\alpha^2\over 2}
    q_\alpha^2\right]\, ,
   \end{equation}
and a linear coupling between the two
    \begin{equation}
   H_{\rm int}= \sum _\alpha\ \xi_\alpha q_\alpha q_0
 \,  .
   \end{equation}
 Formally, the total Hamiltonian 
 $H=H_0 + H_e +H_{\rm int}$ can be written in terms of its normal 
 mode coordinates $X_\nu$ and $P_\nu$:
  \begin{equation}
   H= \sum _\nu\left[ {P_\nu^2\over 2m} + {m\omega_\nu^2\over 2}
    X_\nu^2\right]\, ,
   \end{equation}
and we will consider a situation in which the initial (before the pulse)
wave function corresponds to all the modes in the ground state:
\begin{equation}
\Psi_0= \prod_\nu \left(\omega_\nu \over \pi \hbar\right )^{1/4}e^{-\omega_\nu
    X_\nu^2/2\hbar}.
 \end{equation}   
 
 At $t=0$ a pulse is applied to the (original) oscillator, the wave function
 immediately after the pulse given by:
 \begin{eqnarray}
\Psi_0(t=0^+)&=&e^{i\lambda q_0^2} \Psi_0 \\
&=& e^{i\lambda \sum_{\mu \nu }U_{0\mu}U_{0\nu}X_{\mu}X_{\nu} }\Psi_0\, ,
 \end{eqnarray} 
 where $U_{\mu\nu}$ is the matrix transforming from the original (uncoupled)
 modes to the coupled system ($q_0=\sum _\nu U_{0\nu}X_{\nu}$). 
 
 As in 
 previous sections, we are interested in the fluctuations of the variance 
 of $q_0$, given in this case by
\begin{equation}
\langle q_0^2(t) \rangle =  \sum_{\mu\nu }U_{0\mu}U_{0\nu}
\langle X_{\mu}X_{\nu}\rangle(t) \, ,
\end{equation}  
and that we will compute by solving the equation of motion obeyed by
the correlations $\langle X_{\mu}X_{\nu}\rangle(t)$. 
Since $X_\mu $ and $X_\nu$ correspond to harmonic coordinates, 
using the quantum mechanical commutation relations we compute the 
equations of motion to be:
\begin{eqnarray*}
&&{d\over dt} \langle X_{\mu}X_{\nu}\rangle
={1\over m} \langle (P_{\mu}X_{\nu}+P_{\nu}X_{\mu})\rangle
\\
&&{d^2\over dt^2} \langle X_{\mu}X_{\nu}\rangle
=-(\omega_\mu^2+\omega_\nu^2)\langle X_{\mu}X_{\nu}\rangle
+{2\over m^2}
 \langle P_{\mu}P_{\nu}\rangle\\
&& {d\over dt} \langle P_{\mu}P_{\nu}\rangle
={-m} \left(\omega_\mu^2 \langle X_{\mu}P_{\nu}\rangle 
+\omega_\nu^2\langle X_{\nu}P_{\mu}\rangle \right)\\
&&{d^2\over dt^2} \langle P_{\mu}P_{\nu}\rangle
= -(\omega_\mu^2+\omega_\nu^2)\langle P_{\mu}P_{\nu}\rangle
+ 2m^2 \omega_\mu^2\omega_\nu^2 
\langle X_{\mu}X_{\nu}\rangle \; .
\end{eqnarray*}

Note that the
above equations are identical to those of classical harmonic oscillators
for the quantities $X_{\mu}(t)X_{\nu}(t)$ etc.,
with initial conditions given by the values of the correlations evaluated 
for  the quantum wave function:

\begin{eqnarray*}
&&\langle X_{\mu}X_{\nu}\rangle (0^+)=\delta_{\mu\nu}
 {\hbar\over 2m\omega_\mu},
\\
&&\langle P_{\mu}P_{\nu}\rangle (0^+)=
\delta_{\mu\nu} {\hbar m\omega_\mu\over 2}\\
&&+2\hbar^2 \lambda^2 (1+\delta_{\mu\nu}) {U_{0\mu}\over m\omega_\mu}
{U_{0\nu}\over m\omega_\nu} q_0^2
\\
&&\langle (X_{\mu}P_{\nu}+ P_{\nu}X_{\mu})\rangle (0^+)= 4\lambda 
\hbar U_{0\mu} U_{0\nu} {\hbar\over 2m}(
{1\over \omega_\mu}+ 
{1\over \omega_\nu}) 
\end{eqnarray*}
with $q_0^2\equiv\langle q_0^2(0^-)\rangle
 =\sum_\alpha {\hbar U_{0\alpha}^2/2 m\omega_\alpha}
$.

Collecting the above equations we obtain
\begin{equation}
\langle q_0^2(t) \rangle =q_0^2 \left\{1+ 4\lambda^2 S^2(t) +{\lambda\over q_0^2}
C(t)S(t) \right \},
\end{equation} 
with 
\begin{equation}
S(t) = \sum_\mu {\hbar U_{0\mu}^2\over m \omega_\mu} \sin \omega_\mu t
\,\,
C(t)  = \sum_\mu {\hbar U_{0\mu}^2\over m \omega_\mu} \cos \omega_\mu t.
\nonumber\end{equation}

All the information of the evolution of the variance is contained in the 
function $J(\omega)$, the physical interpretation of which is that of a local density
of states of the oscillator, defined as 
\begin{equation}
J(\omega) =\sum _\mu {\hbar U_{0\mu}^2\over m \omega_\mu} \delta (\omega - \omega_\mu),
\end{equation}
from which
\begin{equation}
S(t)=\int d\omega J(\omega) \sin\omega t , \;\; 
C(t)=\int d\omega J(\omega) \cos\omega t.
\label{sincos}
\end{equation}

Note that $J(\omega)$ is a sum over delta functions, giving rise to a superposition of 
oscillations with the frequencies $\omega_\nu$ for both $S(t)$ and $C(t)$. In the limit of
an infinite system, and when the modes are spatially extended over all space
$J(\omega)$ becomes a continuous
function. In that case the oscillatory behavior acquires a damped component,
the detailed 
time dependence being given by the frequency spectrum of $J(\omega)$. 
A lorenzian shape for $J(\omega)$ will give an exponentially  damped oscillation for both $S(t)$
and $C(t)$. As an illustration of this point we consider a model for which $J(\omega)$ can be 
computed explicitly -- see the classical analysis in Lamb [1900]
Komech [1995].
Consider a one-dimensional  string coupled to our oscillator.
The string is described by a ``transverse" displacement $u(x,t)$. The
 classical equations
of motion of the system are
\begin{eqnarray}
u_{tt}(x,t)&=&c^2  u_{xx}(x,t)\nonumber\\
M d^2{q_0}(t)/dt^2&=&-V{q_0}(t)+T[u_x(0+,t)
- u_x(0-,t)]\nonumber\\
q_0(t)&=&u(0,t).
\label{model}
\end{eqnarray}
 The normal modes consist of even and odd (in $x$) solutions.
 The odd solutions do not involve $q_0$ and are of the form
 $u_{q,o}(x,t) = e^{icqt}\sin qx$, whereas the even solutions are of
 the form
$  u_{q,e}(x,t)=e^{icqt}\cos (q|x| +\delta_q),
 $
  with $\delta_q$ a phase shift (to be found).
  The wave vectors $q$ label the normal modes, and play the role of
  the index $\mu$ in the above discussion: $\omega_\mu = cq$,
  and $U_{\mu0}^2=\cos^2(\delta_q)$ (up to a normalization constant) in the present case. 
  Substituting this expression in 
 (\ref{model}) we obtain (${\omega_0^2=V/M}$)
  \begin{equation}
  \tan \delta_q ={Mc\over 2T}{(\omega_0^2-\omega_q^2)\over\omega_q},
  \end{equation}
from which  $U_{\mu0}^2= \cos^2\delta_q$ is given by
\begin{equation}
U_{\mu0}^2= {\alpha^2\omega_q^2\over \alpha^2\omega_q^2 + ( 
\omega_q^2
-\omega_0^2)^2}
\equiv U_q^2,
\end{equation}
where we have defined
$\alpha= 2T/Mc$. Note that $U_q$ represents the transformation matrix that has 
to be normalized and since the frequencies form a continuum we normalize $U_q(\omega_q)$
to its integral over $\omega_q$. Omitting
 the index $q$ in $\omega_q$, we obtain
\begin{equation}
U(\omega)={2\alpha\over \pi} {\omega^2\over  \alpha^2\omega^2 + ( 
\omega^2
-\omega_0^2)^2}={m\omega \over \hbar}J(\omega).
\label{finalj}
\end{equation}

Substituting (\ref{finalj}) in (\ref{sincos}) we obtain 

\begin{equation}
 S(t) ={\hbar \over m\omega_0} e^{-\Gamma t} \sin \Omega _0 t,\,\,
C(t) ={\hbar \over m\omega_0} e^{-\Gamma t} \cos \Omega _0 t,
 \end{equation}
 with
 \begin{eqnarray}
 \Omega_0& =& \omega_0\left(1 +\left[\alpha/\omega_0\right]^2\right)
^{1/4}\cos \delta/2,\\
\Gamma& =&\omega_0\left(1 +\left[\alpha/\omega_0\right]^2\right)^{1/4}
\sin \delta/2,
\end{eqnarray}
where $\delta = \tan ^{-1} \alpha / \omega_0$.

In the 
realistic limit $\alpha \ll \omega_0$ which corresponds to a ``weak" coupling
to the environment) this expressions take the form: 
$ S(t) \cong(\hbar/(m\omega_0) {\rm exp}(-Tt/Mc)\sin \omega _0 t,
C(t) \cong(\hbar/(m\omega_0) {\rm exp}(-Tt/Mc) \cos \omega _0 t.$

 Note that in this model, and in the limit of weak coupling, the initial variance
 $q_0 ^2$ of the reference oscillator is unchanged due to the coupling to the 
 environment, and is given by $q_0 ^2=\hbar/2m\omega_0$. Our final result for
 this section
 is then

\begin{eqnarray}
\langle q_0^2(t) \rangle \cong q_0^2 \left\{1+ e^{-2(T/Mc) t}\right.\nonumber\\
\left.\left[
\left({2\lambda\hbar\over m \omega_0}\right)
\sin 2\omega_0t
+\left({2\lambda\hbar\over m \omega_0}\right)^2
\sin\omega_0^2t
\right]
 \right\},
\end{eqnarray} 
which reduces simply  to (\ref{nodamp}) in the uncoupled case of $T=0$. 

In summary we have shown in this section that
the coupling to the environment can
be included in general, giving rise to dissipation, 
and that the squeezing effect in the 
presence of dissipation can be computed explicitly for the Lamb model.

Additional details of the analysis here, extensions
to the squeezing of a nonlinear oscillator, and a treatment of the 
quantum measurement issue  will appear in forthcoming publications.

{\bf Acknowledgement:} We would like to thank Roger Brockett for useful
discussions.

\noindent{\bf References}
\begin{description}

\item Brockett, R. and N. Khaneja [1999] On the stochastic control of 
quantum ensembles, preprint.

\item Garret, G.A., A.G. Rojo, A.K. Sood, J.F. Whitaker and R. Merlin [1997]
Vacuum Squeezing of solids: Macroscopic quantum states driven by light
pulses, {\it Science} {\bf 275}, 1638-1640.

\item Hagerty, P. A.M. Bloch and M.I Weinstein [1999] Radiation 
induced instability in interconnected systems, {\it Proc. 37th Conference
on Decisions and Control}, IEEE, 651-656.

\item Keller, J.B. and L.B. Bonilla [1986] Irreversibility
and nonrecurrance, {\it Journal of 
Statistical Physics} {\bf 42} 1115.

\item Komech, A. I. [1995] On the stabilization of
string-oscillator interaction. {\it Russian Journal of
Mathematical Physics} {\bf 3}, 2, 227-247.

\item Lamb, H. [1900] On the peculiarity of the wave-System due to
the free vibrations of a nucleus in an extended medium, {\it
Proceeding of the London Math. Society}, {\bf 32}, 208-211.

\item Lloyd,  S. [1996] Universal quantum simulators, {\it Science}
{\bf 273}, 1073-1078.

\item Negele, J. W.  and  H. Orland [1988] {\it Quantum Many Particle Physics},
 Addison-Wesley, New York.

\item Soffer, A. and Weinstein, M.I. [1999] Resonances, Radiation
Damping and Instability in Hamiltonian Nonlinear Wave Equations,
{\it Invent. Math.} {\bf 136} 9--74.

\item Warren, W., H. Rabitz, and M. Dahleh [1993] Coherent control of 
quantum dynamics: The dream is alive, {\it Science} {\bf 259}, 1581-1589.

\item Ziman,  J. [1972] {\it Principles of The Theory of Solids}, Second Edition,  Cambridge 
University Press.

\end{description}

\end{document}